\documentclass[%
 reprint,
 amsmath,amssymb,
 aps,
floatfix,
]{revtex4-2}

\usepackage{graphicx}
\usepackage{dcolumn}
\usepackage{bm}

\begin{document}

\preprint{APS/123-QED}

\title{Telecom-Band SPDC in AlGaAs-on-Insulator Waveguides}

\author{M.~Placke$^{1,*}$, J. Schlegel$^{2}$, F.~Mann$^{1}$, P. Della Casa$^{2}$, A. Thies$^{2}$, M. Weyers$^{2}$, G. Tr\"{a}nkle$^{2}$, S. Ramelow$^{1, 3}$}

\affiliation{$^1$ Institute for Physics, Humboldt-Universit\"{a}t zu Berlin, Newtonstra\ss e 15, 12489 Berlin, Germany \\
$^2$ Ferdinand-Braun-Institut, Gustav-Kirchhoff-Stra\ss e 4, 12489 Berlin, Germany \\
$^3$ IRIS Adlershof, Humboldt-Universit\"{a}t zu Berlin, Zum Großen Windkanal 2, 12489 Berlin, Germany
}

\date{\today}

\begin{abstract}
Widespread commercial adoption of telecom-band quantum-key-distribution (QKD) will require fully integrated, room-temperature transmitters. Implementing highly efficient spontaneous parametric down-conversion (SPDC) on a platform that offers co-integration of the pump laser has been an outstanding challenge. Here, using such a platform based on AlGaAs-on-insulator waveguides, we report telecom-band SPDC (and second harmonic generation) with exceedingly large efficiencies of 26 GHz generated pairs/mW over a 7 THz bandwidth, which would saturate the usable photon-flux for a 70-channel wavelength-multiplexed QKD-system at merely 1.6 mW of pump laser power. 
\end{abstract}

\maketitle


\section{Introduction}
Integration of nonlinear optics onto chip level not only promises cost-effective, real-world applicability, but comes with its own set of requirements and trade-offs compared to bulk-optics solutions depending on the specific application.
While many nonlinear optical materials offer a variety of different sets of benefits, III-V semiconductors (SCs) stand out as a particularly versatile platform for integrated photonics with remarkable features: For these materials, the very high nonlinearities for both second- and third-order parametric processes are equally noteworthy as is the exceptional quality of heteroepitaxial fabrication techniques and most importantly the possibility for monolithic on-chip integration of pump lasers on the same chip. Moreover, compound compositions can be adapted for transparency from visible to mid-infrared wavelengths and, due to the semiconductor-typical large refractive indices, the design of waveguides, which form the most fundamental ingredient of photonic integration, can access powerful dispersion engineering with great flexibility in the choice of claddings. Accordingly, cladding materials can range all the way from nearly seamless successions of lattice-matched hetero-epitaxial semiconductor layers to suspended waveguide cores surrounded only by air.
With the choice between such extremes, another tradeoff between application-specific performance and versatility of the integrating framework is typically required for III-V integrated nonlinear optics: Evidently, monolithic approaches can access a richer spectrum of established optoelectronic functionalities, while their small refractive index contrasts yield rather small effective nonlinearities and weak dispersion control compared to III-V-on-insulator architectures. With its very large second- and third-order material nonlinearities even compared to other III-V compounds, Al$_x$Ga$_{1-x}$As marks a highly anticipated option for parametric three-wave- and four-wave-mixing - a resource of great interest for both classical and quantum photonic technology \cite{mi13070991, Baboux:23}. Here, monolithic approaches with integrated pump lasers such as Bragg reflection waveguides \cite{Thiel:23, Yan:22} coexist with highly confining AlGaAs-on-insulator waveguides \cite{Kuyken:20, Xie:20}. Whilst the former use second-order nonlinearities to produce photon pairs via spontaneous parametric down conversion (SPDC), nonlinear optics in AlGaAs-on-insulator devices has so far typically exploited third-order nonlinearities. Nevertheless, as the fabrication quality and number of available photonic building blocks in the on-insulator architecture is steadily advancing \cite{10.1063/5.0098984, doi:10.1126/science.aan8083}, accessing both second- and third-order wavelength conversion makes AlGaAs-on-insulator waveguides a more and more compelling alternative to the most popular competitors such as thin-film lithium niobate waveguides \cite{Zhu:21, Rao:19} and high-Q Si$_3$N$_4$ resonators \cite{10.1063/5.0057881}. 

In this study, we present highly efficient $\chi^{(2)}$-conversion from and to telecom wavelength ranges in such waveguides. Our waveguide (dispersion) design and (expected) efficiency calculations adapts our previous simulation results \cite{Placke:20} that targeted C-band phasematching. Whilst efficiencies on the same order of magnitude as these simulation figures were successfully demonstrated in GaAs-on-insulator for second harmonic generation (SHG) from 2 to 1 \textmu m \cite{Stanton:20}, high losses towards shorter second harmonic wavelengths prevented similar performance in AlGaAs-on-insulator waveguides pumped from the telecom range \cite{mi11020229, May:19}. Consequently, SPDC has yet remained unreported since most single-photon-sensitive detectors require light emission to stay below or near telecom wavelengths.

\begin{figure*}[htp]
\centering
\includegraphics[width=\linewidth]{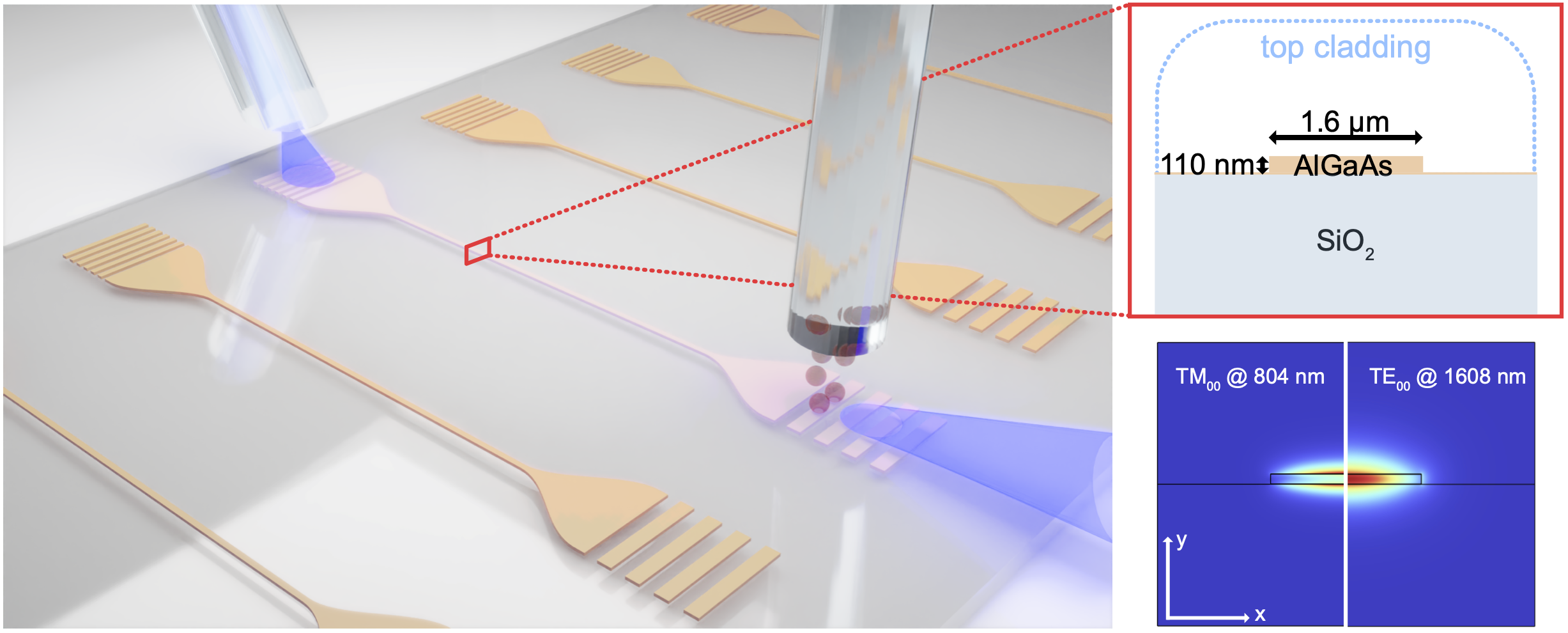}
\caption{Left panel: the semiconductor layer is lithographically structured to produce fiber size-matched grating couplers for first and second harmonic bands, while the tapered dimensions of the conversion section tune the phase matching to the desired range of telecom wavelengths. Upper right: After precharacterization at the wafer-level, the AlGaAs rib-waveguides residing on SiO$_2$ bottom cladding may receive additional top cladding deposition for modification of waveguiding and phasematching properties. Lower right: the nonlinear conversion efficiencies are derived from overlap integrals of the fundamental (TE- and TM-) modes as commonly obtained from simulation with FEM solvers.}
\label{fig:render}
\end{figure*}

\section{Methods}

Our waveguide fabrication structures the 110 nm thick III-V SC stack (residing on a SiO$_2$ bottom cladding, see Figure \ref{fig:render}) to make 1.6 \textmu m wide rib waveguide cores with a shallow residual slab that has negligible influence on the waveguide's Eigenmodes. In this thin-film geometry, the large amplitude-evanescence along the vertical (fast) waveguide axis is balanced to compensate for the (bulk) material dispersion between first and second harmonic frequency. Accordingly, the transverse magnetic (TM) fundamental mode at second harmonic frequency propagates with the same effective phase index as the TE$_{00}$-mode in the telecom band such that phasematching can be achieved for SHG and degenerate SPDC by finetuning the thickness and width of the waveguide. For this and to estimate the resulting efficiencies, we obtained the specific waveguide dispersion for and mode overlaps between the nonlinearly coupled TM- and TE-mode from finite element method (FEM) simulations. Assuming a bulk material nonlinearity of $d_{36} = 100$ pm/V for Al$_{0.2}$Ga$_{0.8}$As \cite{Shoji:97} the simulated nonlinear efficiencies amount to \begin{equation}
    \tilde{\eta}^{\textrm{sim}}_{\textrm{shg}} = \frac{P_{\textrm{shg}}}{P^2_{\textrm{p}} \ L^2} = \frac{8 \pi^2 \Gamma_d }{c_0 \lambda_{\textrm{p}}^2 \epsilon_0 n_{\textrm{eff}}^3}\approx 189 \ \%/\textrm{W}/\textrm{mm}^2
\end{equation}
for SHG \cite{risk_gosnell_nurmikko_2003} and
\begin{equation}
\tilde{\eta}^{\textrm{sim}}_{\textrm{spdc}} = \tilde{\eta}^{\textrm{sim}}_{\textrm{shg}} \ h \nu_{\textrm{si}} \ \Delta \nu^{\textrm{sim}}_{\textrm{si}} \ \sqrt{L} \approx 1.98 \cdot 10^{-6} \cdot \textrm{mm}^{-3/2}    
\end{equation}
for degenerate SPDC \cite{Fiorentino:07}. The nonlinear overlap is approximated by the main vectorial components of both FH and SH Eigenmodes
\begin{equation}
   \Gamma_d \approx \frac{\int \ d_{36}(x, y) \ E^2_{\textrm{FH,x}}(x, y) \ E_{\textrm{SH,y}}^*(x, y) \ dx \ dy}{\int \left| \vec{E}_{\textrm{FH}} \right|^2 \ dx \ dy \ \ (\int \left|\vec{E}_{\textrm{SH}}\right|^2 \ dx \ dy)^{1/2}} \quad .
\end{equation}
Further, $\lambda_{\textrm{p}}$ gives the wavelength of the telecom pump laser and $n_{\textrm{eff}}$ denotes the effective refractive index of the waveguide modes at zero detuning from phase matching.

The bandwidth-length-scaling can in first order be derived solely from the group velocity dispersion at the signal-idler-degeneracy point and yields $\Delta \nu^{\textrm{sim}}_{\textrm{si}} = 8.5 \ \textrm{THz} \cdot$mm$^{1/2}/\sqrt{L}$. Hence, considering the specific length $L_{\textrm{nl}} \approx 1.44$ mm of the $\chi^{(2)}$-conversion section of our waveguide device, signal-idler photons with a telecom band-spanning $\Delta \nu^{\textrm{sim}}_{\textrm{si}} \approx $ 7.1 THz bandwidth are predicted from the modelling. 

For efficient coupling to single mode fibers, the rib waveguides in our design  are tapered to 9 \textmu m width and terminated with single etch-step grating-couplers on either side of the nonlinear conversion section. By virtue of these grating-based coupling interfaces, moderate mode-matching (with insertion losses down to 4 dB) between the highly confining waveguides and the bulky fibers merely requires lithographic structuring with lateral feature sizes $\geq$ 250 nm. Consequently, the waveguide patterning can be performed with a variety of photo- and electron beam lithography processes out of which the most suitable recipes for high surface qualities can be chosen.\\

\section{Experimental results}
\subsection{Reducing propagation losses}
\begin{figure}[htp]
\centering
\includegraphics[width=\linewidth]{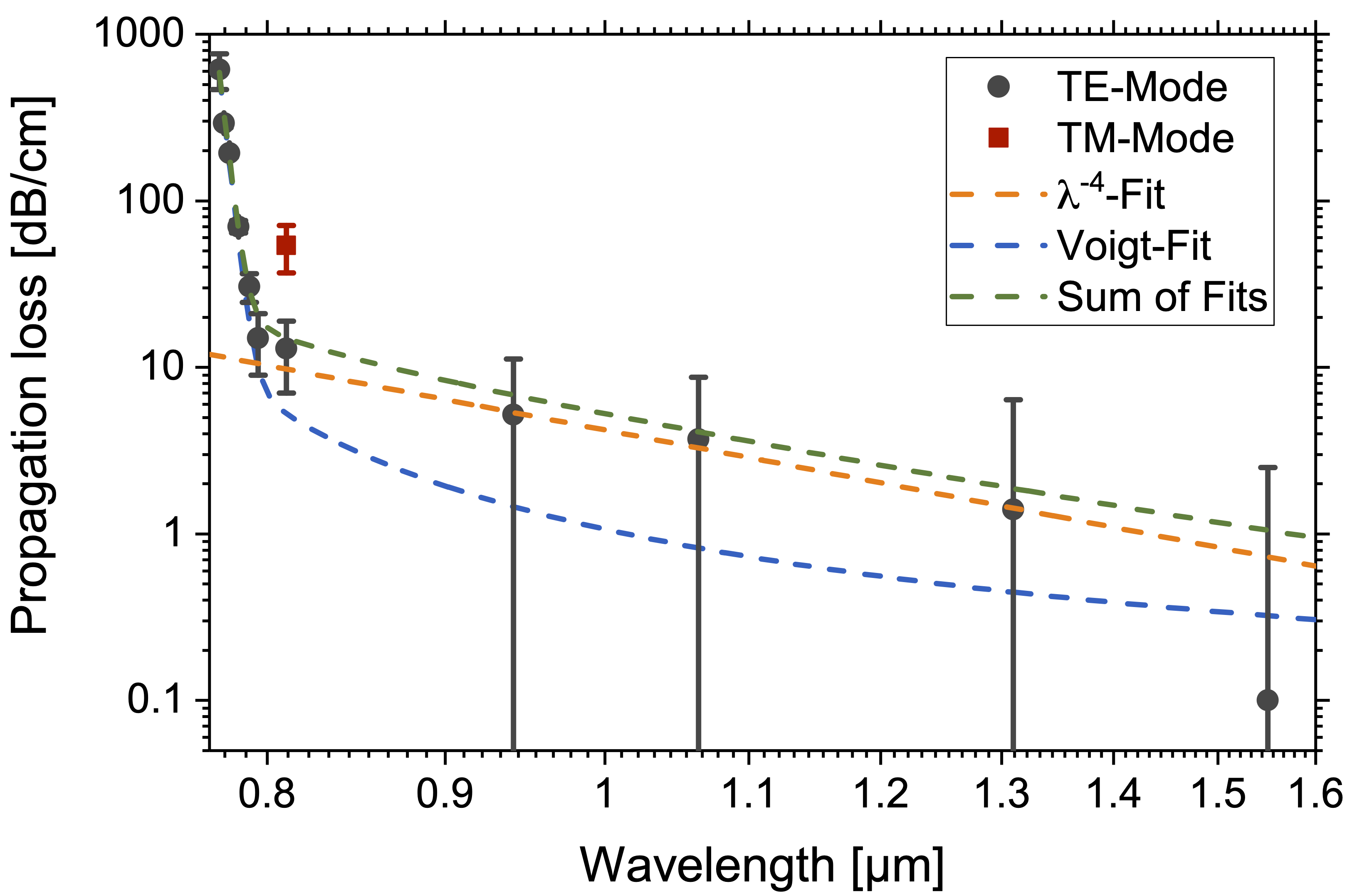}
\caption{Loss spectra before top cladding deposition: A steep onset of absorption losses (blue) beyond the scattering loss-baseline (orange) is found towards shorter wavelengths ($\lambda < 795$ nm).}
\label{fig:cutback}
\end{figure}

In order to achieve efficient SHG and SPDC it is absolutely vital to reduce propagation losses in the waveguide especially for the NIR wavelengths around 800 nm. To pre-characterize and subsequently improve the propagation losses, waveguide sections with a range of different lengths were fabricated ("cut-back structures") which - together with the robustness of the coupling interface - allowed us to measure the mode propagation losses level in a very wide range values (3-300 dB/cm). Afterwards we applied and studied different selective surface treatments e.g. wet chemical passivation and/or top layer deposition to individual chips.
The measurements at the wafer-level and for the bare waveguide cores (without top cladding), resulting in a spectral loss analysis, see Fig. \ref{fig:cutback}, find that parasitic absorption (at energies below the bandgap of bulk Al$_{0.2}$Ga$_{0.8}$As with $\lambda_{\textrm{gap}} \approx 743$ nm) makes operation exceedingly lossy only for wavelengths below 795 nm. Below 795 nm we observe a very steeply increasing contribution from absorption losses, which we model by a Voigt profile (blue dashed). Above 795 nm, the measured TE-mode cutback data coincides well with the (orange dashed) extrapolation that mimics Rayleigh scattering by following a $\lambda^{-4}$-dependence from the four longest-wavelength datapoints. The sum of both fit functions (green dashed) agrees very well with the complete experimental data. A similar behavior with a sudden onset of absorption well before the bandgap has previously been reported in GaAs thinfilm waveguides, albeit for wavelengths below 1 \textmu m \cite{Stanton:20}. In this case, As-As dimers at the reconstructed surface were assumed to produce the parasitic density of states within the bandgap that effectively reduced the waveguide's transparency by more than 13\% compared to the bulk SC. Indeed, first principles electronic structure calculations find the antibonding state of the As-As dimer at the semiconductor-to-oxide interface \cite{10.1116/1.4710513} to provide a significant partial density of states below the conduction band minimum of bulk-GaAs. As these dimers - unlike common defect-related mid gap states with a surface coverage of $<$0.1\% \cite{Kamel:23} - occur with nearly unit cell periodicity, they can give rise to strong surface state (linear  i.e. regardless of the injected optical power) absorption. Further, since (first principles) electronic states predominantly depend on the local binding configuration, the dimers can be expected to produce similar gap states in ternary AlGaAs alloys too. Due to a reduced bond length in InP, its anion dimer's antibonding state lies well above the conduction band minimum. Again, similar electronic characteristics can be expected for the ternary InGaP alloy lattice-matched to AlGaAs such that the InGaP-to-oxide boundary might cause less gap state-induced absorption. Consequently, we have adapted the waveguide core stack to consist of a 100 nm thick AlGaAs in between two lattice matched InGaP layers of 5 nm thickness such that the higher nonlinear susceptibilities of AlGaAs - although both compounds only differ significantly in their third order susceptibilities - can largely be utilized while the surface-state absorption losses at short wavelength may be significantly reduced. The spectral dependence in Fig. \ref{fig:cutback} suggests that for all-optical operation, limitations from unpassivated III-V surfaces can be sidestepped with much smaller sacrifice of transparency bandwidth  ($\Delta_{\textrm{trans}}/\Delta_{\textrm{gap}} \approx  94 \ \% $) in our InGaP/AlGaAs/InGaP waveguide stacks compared to previous loss observations in (Al)GaAs-on-insulator. In consequence, our III-V SC waveguide design accesses the largest parametric nonlinearities of all state-of-the-art (solid state) materials for integrated photonics with a transparency that is compatible with the telecom L-band and it's range of second harmonic frequencies. Although  InGaP-on-insulator provides similar second-order nonlinearities \cite{Zhao:22}, third-order interactions are limited to much smaller efficiencies.
\begin{table}[htbp]
\centering
\caption{\bf Comparison of various top claddings}
\begin{tabular}{cccc}
\hline
material & thickness & \small{phase-matched} $\lambda_{\textrm{p}}$& loss at $\lambda_{\textrm{p}}$/2\\ 
\hline
 \small{no top-claddin}g& - & $<$1550 nm & $>$300 dB/cm\\
\small{IC-PECVD SiO$_2$} & 100 nm& 1573 nm& 30 dB/cm\\
 \small{IC-PECVD SiO$_2$}& 200 nm& 1579 nm&30 dB/cm\\
\small{PECVD Si$_3$N$_4$}  & 50  nm& 1599 nm & 70 dB/cm\\
 \small{PECVD Si$_3$N$_4$}& 100 nm& 1627 nm&70 dB/cm\\
\small{Spin-on AZ1518} & 1.6 \textmu m& 1608 nm& 14 dB/cm\\
\hline
\end{tabular}
  \label{tab:Redshift}
\end{table}
More importantly, realizing low-loss propagation down to 795 nm now enables highly efficient conversion between the telecom L-band (1565-1625 nm) and its corresponding second harmonic range. Here, we benefit from the vastly uneven sensitivitiy of the nonlinear conversion section and the coupling section to additional or modified cladding layers: Whilst the coupling efficiency remains largely unaffected (or even improved due to reduced Fresnel reflection losses), the point of phase matching is efficiently shifted (to the red), when the short-$\lambda$ TM-mode's evanescent field samples a larger refractive index top cladding.
\begin{figure}[htp]
\centering
\includegraphics[width=\linewidth]{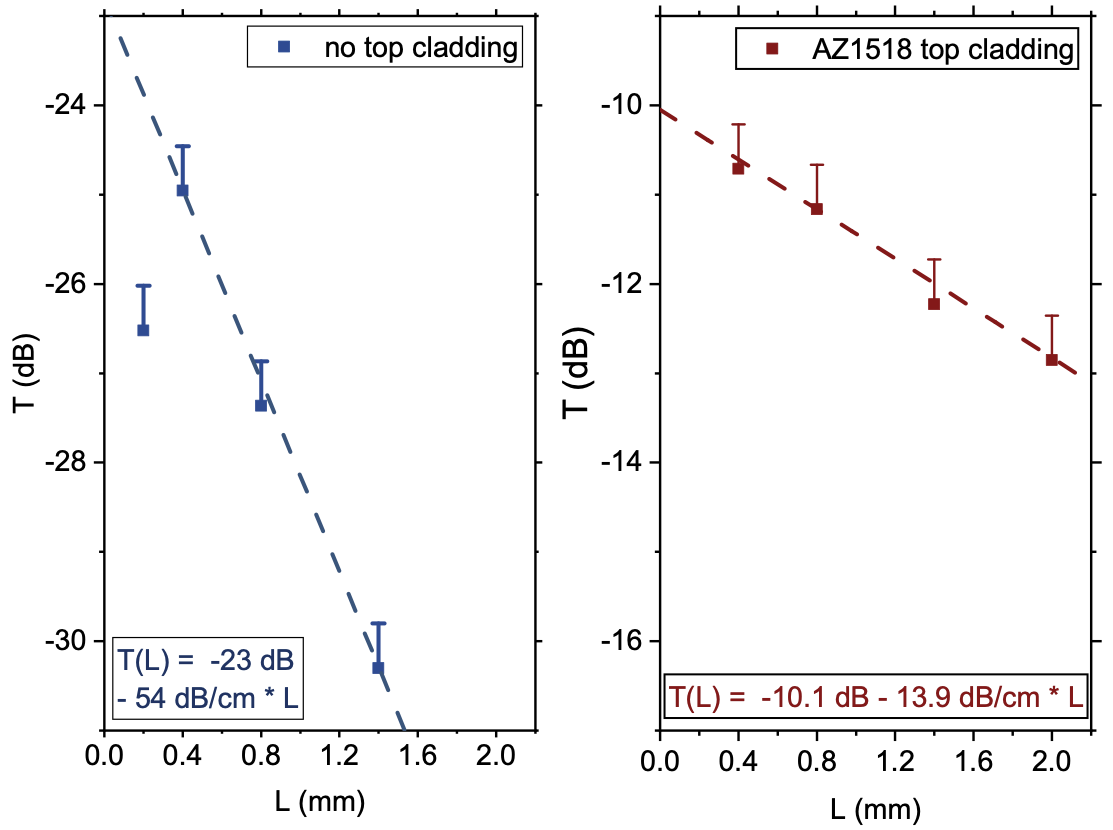}
\caption{Waveguiding before and after AZ1518 deposition: fourfold decreased propagation losses are found for the SH-mode at the post-deposition (SHG-signal) phasematching wavelength of 804 nm.}
\label{fig:newcutback}
\end{figure}
Table \ref{tab:Redshift} shows the central wavelength of SHG detuning curves obtained for various choices of top cladding deposited on individual chips of the pre-characterized (and top cladding-free) wafer. Apart from their thicknesses, these layers vary in terms of their structural and optical properties due to different deposition techniques and material compositons. Here, a comparison of cutback data with Fig. \ref{fig:cutback} serves as a sensitive probe for the optical quality of each additional deposition. As Figure \ref{fig:newcutback} shows, we find strongly reduced propagation losses of the short-$\lambda$ TM-mode for the spin-coated photoresist top cladding (AZ1518). This improvement may result from the AZ1518 embedding previous resist leftovers that contaminated the bare SC surface after lithographic structuring. Potentially the spin-coated cladding might also be favorable over the thinner PECVD and sputtered layers due to near-zero overlap of the waveguide modes with the topmost (air) interface.\\
\subsection{SHG}
\begin{figure}[htp]
\centering
\includegraphics[width=\linewidth]{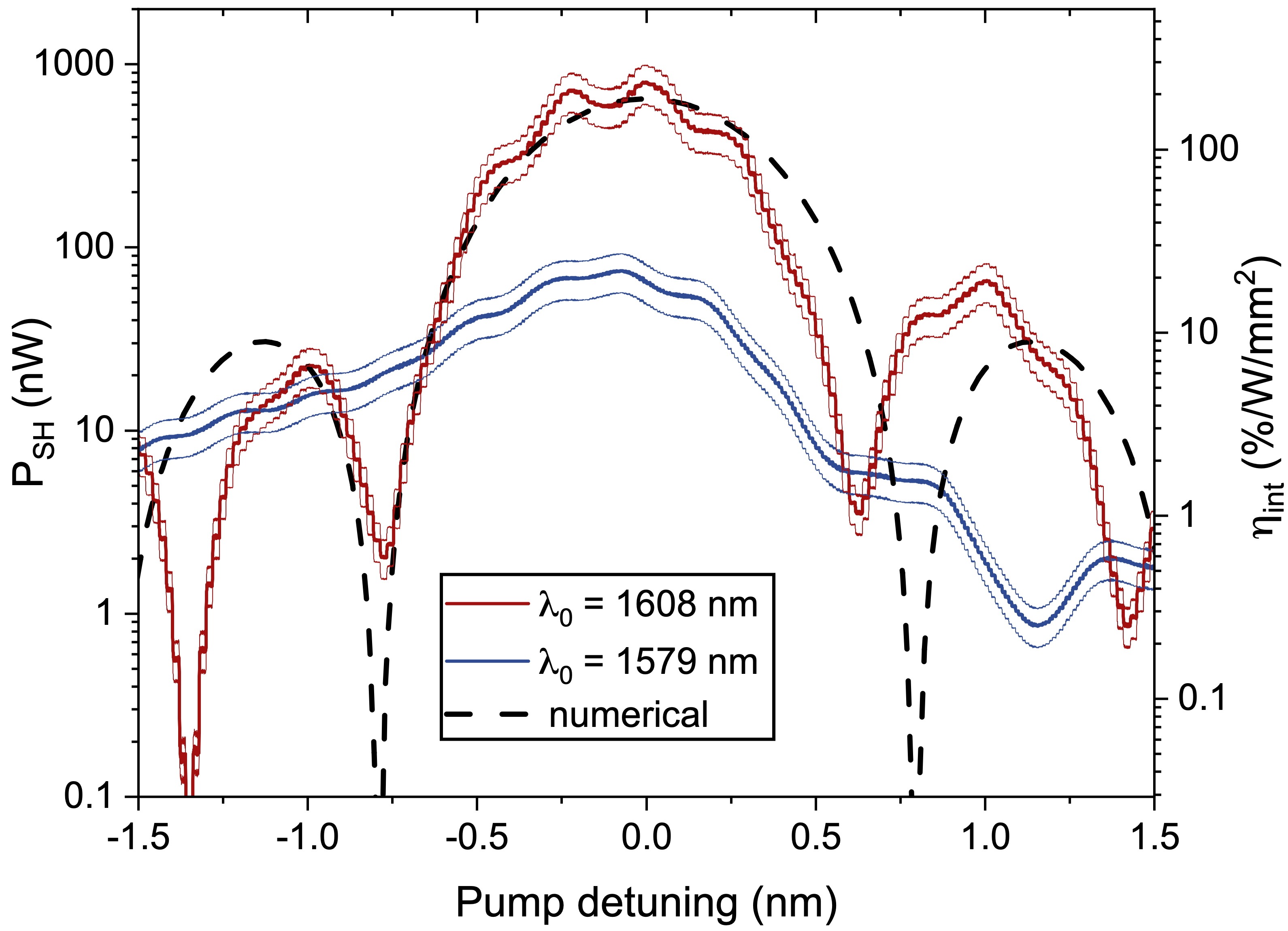}
\caption{Second harmonic detuning for AZ1518 vs. SiO$_2$ top cladding: A tenfold (SHG) efficiency improvement is obtained for a phasematching (SHG-pump) wavelength redshifted to 1608 nm by 1.6 \textmu m of the spin-on resist compared to a (200 nm) PECVD SiO$_2$-coated chip exhibiting phasematching at 1579 nm.}
\label{fig:detune}
\end{figure}

Together with the phasematching redshift to L-band pump wavelengths $\lambda^{\textrm{AZ}}_{\textrm{p}} = 1608$ nm, the spin coating-cladded waveguides yield a strongly improved SHG efficiency as compared to the SiO$_2$-cladded chips with $\lambda^{\textrm{Ox}}_{\textrm{p}} = 1580$ nm as seen in Figure \ref{fig:detune}. Taking the coupling efficiencies of both pump and signal modes into account, we reconstruct the internal SHG efficiency normalized to the 1.44 mm length of the nonlinear conversion section to be $\tilde{\eta}^{\textrm{int}}_{\textrm{shg}} \approx $ 233 \%/ W/ mm$^2$. This value marks a tenfold (fivefold) improvement in comparison with previously obtained values in AlGaAs (LNOI) \cite{mi11020229}
\cite{Rao:19}
waveguides at telecom pump wavelengths. Furthermore, the total transmission through the (comparatively short-length) conversion section is now only weakly diminished by the remaining -14 dB/cm of TM propagation losses such that the reconstructed internal efficiency (see Fig. \ref{fig:detune}) is in good agreement with simulated values for lossless waveguides, which it even mildly exceeds. 
\subsection{SPDC}
\begin{figure}[htp]
\centering
\includegraphics[width=\linewidth]{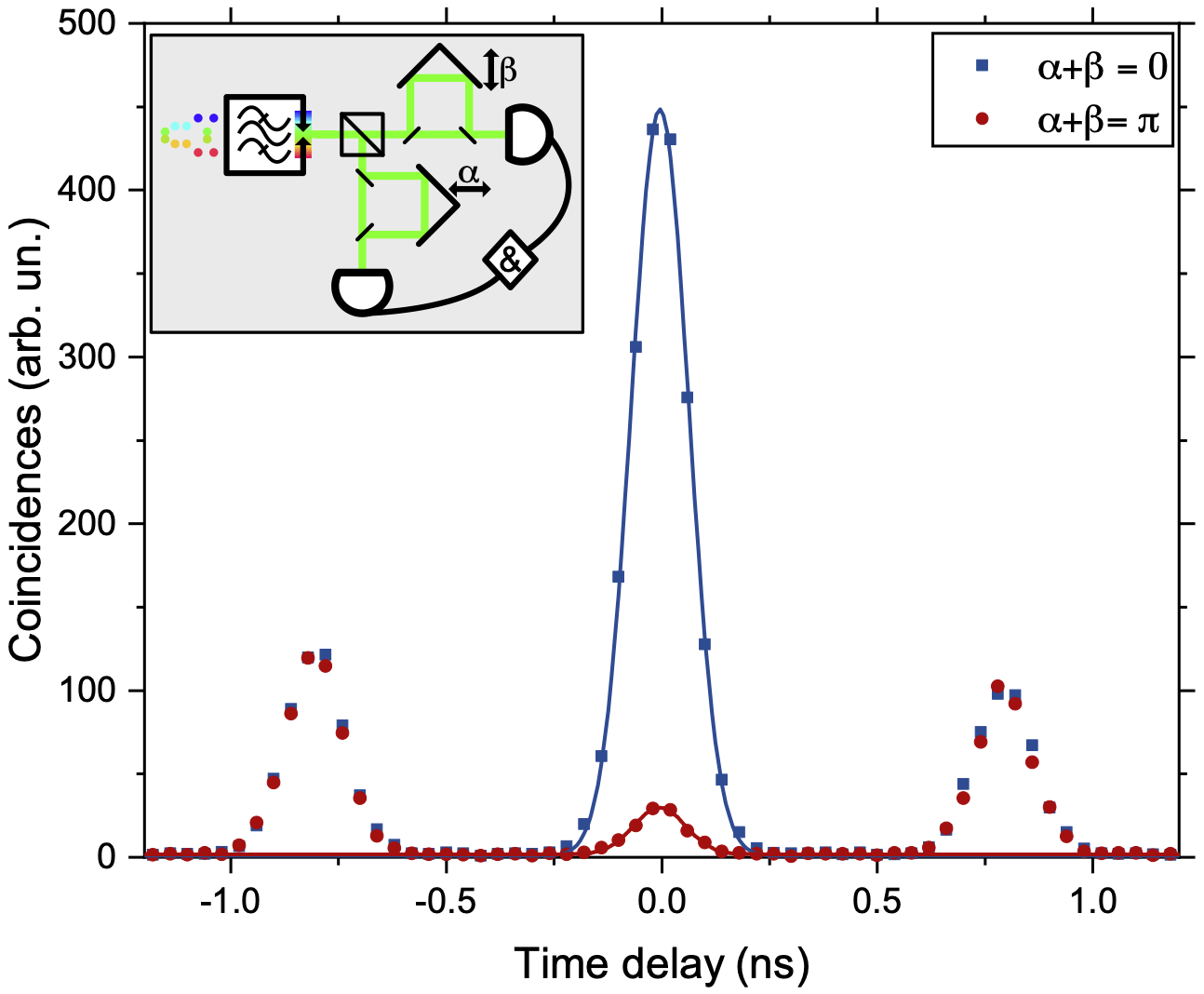}
\caption{For Franson interference the signal-idler emission was narrowly bandpass-filtered and passed through asymmetric Mach-Zehnder interferometers (AMZI) in each (post-beamsplitter) arm, see inset. The autocorrelation shows a visibility of $V = 88.3 \pm 1.2$ \% at zero time delay between con- and destructive AMZI phase settings (main panel). The zero-delay detection events - corresponding to signal-idler photons both travelling either the short or the long AMZI path - exhibit a two-photon interference that violates the commonly applied CHSH-limit of $V \leq 70.7$\% for classical correlations.}
\label{fig:Franson}
\end{figure}

Once the SHG phase-matching point is known, it is straightforward to find SPDC by setting the pump laser to the wavelength of peak SH power and injecting it from the short-$\lambda$ coupler. On the other end, signal and idler photons are collected with an SMF-28 single mode fiber followed by a long pass filter stage with 82 \% transmission at $\lambda_{\textrm{si}}$. Subsequently a (nominally) 50/50 beamsplitter splits the emission into two arms between which coincidences are recorded by two SNSPD detectors with a detection efficiency of $\eta_{\textrm{det}} \approx 85$ \% at $\lambda_{\textrm{si}}$. Assuming equal detection efficiencies of both detectors, we find an effective splitting of 42/58\% between both arms.
\begin{figure}[htp]
\centering
\includegraphics[width=\linewidth]{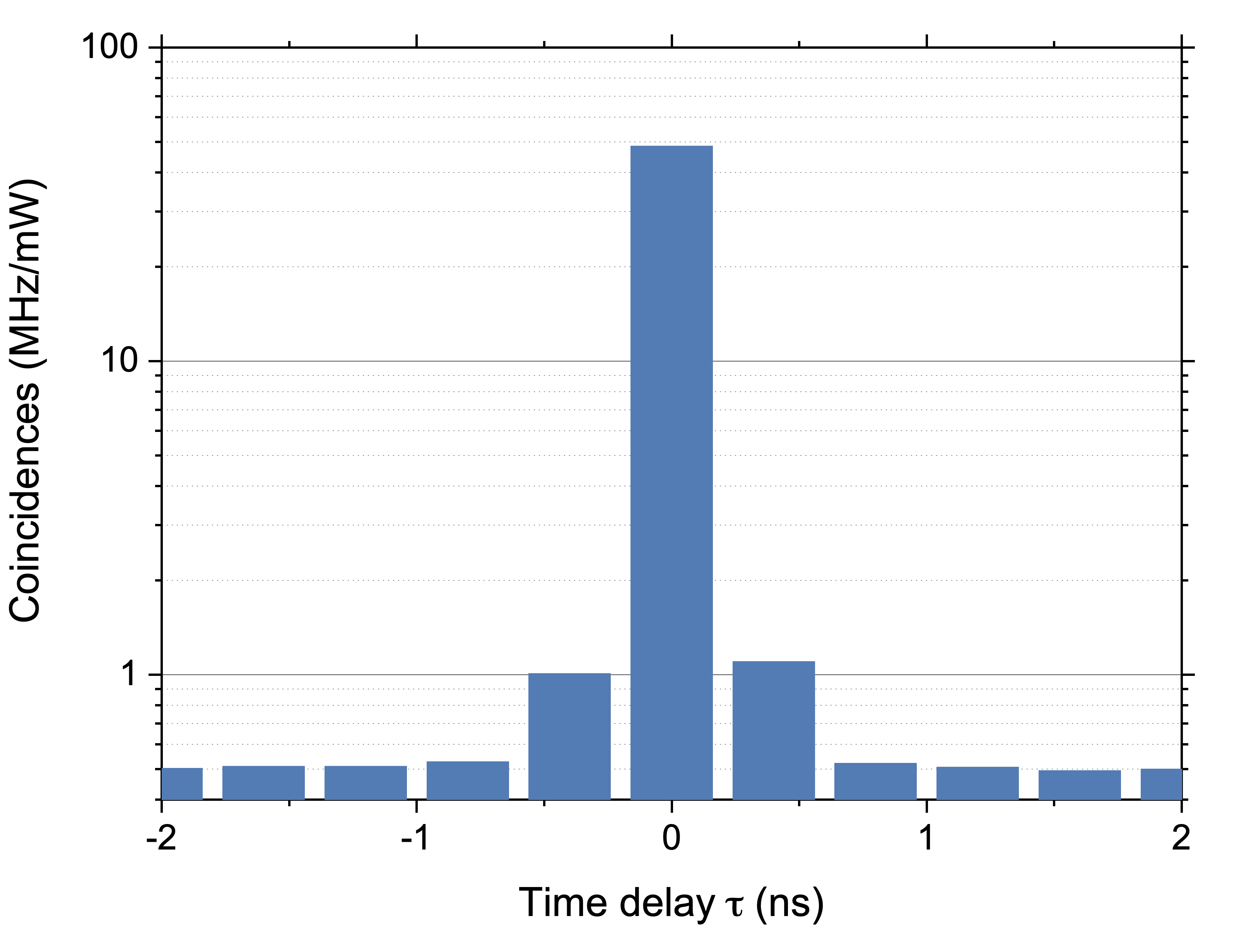}
\caption{Signal-idler autocorrelation: The normalized degree of second-order coherence is found to be $g^{(2)}(0) = 220.9 \pm 1.3$. demonstrating the strong temporal corrrelations of the photon pair emission process far beyond the degree of photon bunching in (classical) thermal light that is characterized by $g^{(2)}(0) = 2$.}
\label{fig:g2}
\end{figure}
Figure \ref{fig:g2} shows the measured biphoton autocorrelation. For our broadband signal-idler emission, the biphoton coherence time falls well below the timing resolution of our coincidence detection ($ T_{\textrm{res}} \approx 150 $ ps) such that little information about the signal-idler spectrum can be inferred from the autocorrelation function's temporal shape. However, the normalized degree of second-order coherence  $g^{(2)}(0) = 220.9 \pm 1.3$ verifies the strong temporal correlations of the photon-pair emission for a measured coincidence rate of $R_{\textrm{cc}}= 48.1 \pm 0.5$ kHz achieved by pumping the waveguide with 3.8 (1) \textmu W launched (coupled) power. The corresponding single counts rates are $S_1 = 0.95 \pm 0.02$ and $S_2 = 1.35 \pm 0.02$ MHz. When corrected for losses due to pump filtering, probabilistic splitting (with ratio $S_1:S_2$), detector efficiency, and biphoton coupling loss, these measured values correspond to a heralding efficiency of $\eta_{\textrm{herald}} = 13.0 \pm 0.2 $ \% and a normalized pair detection rate of 48 MHz/mW. Taking pump coupling loss into account, an internal SPDC generation efficiency of $\eta^{\textrm{int}}_{\textrm{spdc}} = \frac{S_1 S_2}{R_{\textrm{cc}} \  N^{\textrm{int}}_{\textrm{p}} }  \approx 6.5 \cdot 10^{-6}$ (pairs per pump photon) corresponding to a power-normalized generation rate of $\tilde{R}^{\textrm{int}}_{\textrm{spdc}} \approx 2.6 \cdot 10^{10} \frac{\textrm{pairs}}{\textrm{s}\cdot  \textrm{mW}}$ is reconstructed with $N^{\textrm{int}}_{\textrm{p}}$ giving the number of waveguide-internal pump photons. This roughly agrees with the theoretically predicted value of $\eta^{\textrm{sim}}_{\textrm{spdc}} = 3.4 \cdot 10^{-6}$ (pairs per pump photon). Again, the experimental values exceed the theoretical predicitons obtained for a bulk nonlinear susceptibility of $d_{36} = 100$  pm/V. As material parameters differ rather strongly within the literature, our assumption of $d_{36}$ might underestimate the real nonlinearity. Further, the grating coupling efficiencies of the device under test are only inferred from cutback measurements on neighboring waveguides on the same chip. Hence, deviations from the mean grating coupler performance in the nonlinear device under test will distort the reconstructed internal conversion efficiencies.\\
Finally, to demonstrate the high quality of entanglement generated by the source we conducted a Franson interference experiment by narrowly filtering ($\Delta \lambda_{\textrm{FI}} \approx 200 \ \textrm{pm}$) the signal-idler emission at the degeneracy point ($\lambda_{\textrm{si}} \approx 1608.3$ nm) and inserting  asymmetric Mach-Zehnder interferometers in each arm between the beamsplitter and each detector. Fig. \ref{fig:Franson} shows the variation of correlated counts between the two detectors for phase settings corresponding to constructive and destructive interference between coincident detection events (at $\tau = 0$). From the fitted Gaussian amplitudes, we find a visibility of 88.3 $\pm$ 1.2 \%, which exceeds the CHSH-limit of 70.7 \% for classical correlations by fourteen standard deviations and thereby demonstrates the non-classicality of the signal-idler emission. Nevertheless, to genuinely verify (time-energy) entanglement in the Franson interferometer, the postselection loophole must be accounted for, which imposes a more rigorous visibility limit beyond 94.6\%. To achieve such high visibilities, narrower filtering may become necessary to restore indistinguishability between short-short and long-long coincidences, i.e. raise the  (filtered) coherence length beyond the difference of short-long path length differences. Further improvements of fiber coupling stability and efficiency may in turn become necessary to lower the accidental background and mitigate multi-photon emission at a given (adjusted) coincidence
count rate. However, such optimization was beyond the scope of this study.
\section{Discussion}
\begin{figure}[htp]
\centering
\includegraphics[width=\linewidth]{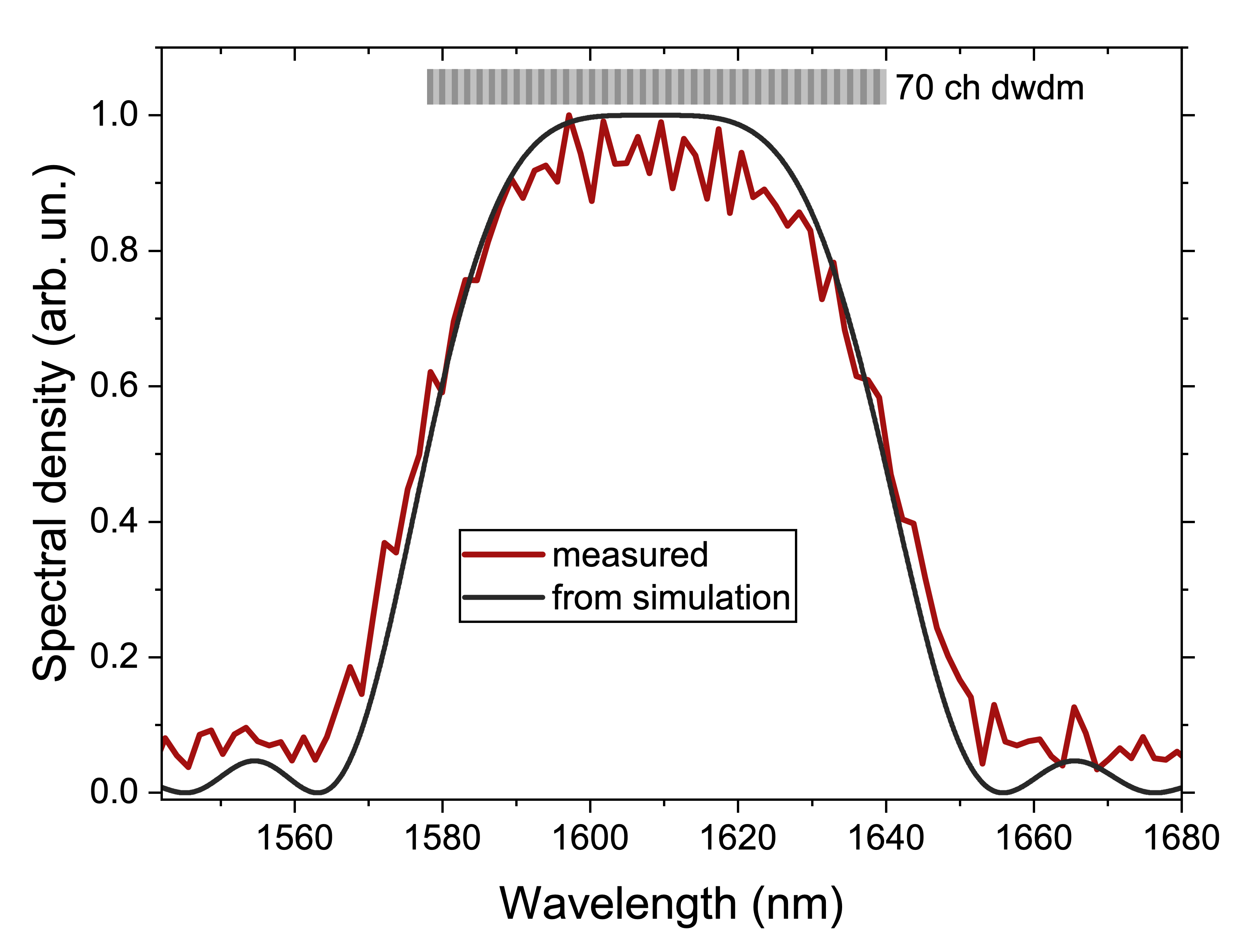}
\caption{Signal-idler spectrum at the point of degenerate phase-matching: a bandwidth of $\Delta \nu_{\textrm{si}} \approx 7.1 \ \textrm{THz} \Leftrightarrow 61$ nm at full width half maximum is found for the 1.44 mm long waveguide in precise agreement with simulation results. The biphoton emission spans over 70 (100 GHz) DWDM channels (grey bars).}
\label{fig:spectrum}
\end{figure}

The demonstrated, exceptionally strong, effective $\chi^{(2)}$-nonlinearity significantly lowers the pump power demands for using this source of entangled photon pairs for quantum communicatioin tasks and thereby improves the scalability of corresponding quantum key distribution schemes. To estimate the internal pump power that is required for an exemplary key-rate optimized telecom-band multiplexed communication scheme, we calculate
\begin{equation}
    P^{\textrm{int}}_{\textrm{p}} = \frac{\mu \ \Delta \nu_{\textrm{si}} }{T_{\textrm{res}} \Delta \nu_{\textrm{ch}}} \ \cdot \ (\tilde{R}^{\textrm{int}}_{\textrm{spdc}} [\frac{\textrm{pairs}}{\textrm{s} \cdot \textrm{mW}} ])^{-1} \approx 1.6 \ \textrm{mW}
\end{equation}
Here, a tradeoff between low multiphoton emission events and high key-rates is made by limiting the signal-idler generation per resolvable time bin of 150 ps to $\mu = 0.05$. Also, a 100 GHz channel spacing with $\Delta \nu_{\textrm{ch}}$ = 54 GHz (FWHM) passbands has been assumed. The signal-idler emission spectrum was obtained with a sensitive grating spectrometer, see Fig. \ref{fig:spectrum} and the bandwidth was found to be $\Delta \nu_{\textrm{si}} \approx 7.1 \ \textrm{THz} \leftrightarrow 61$ nm in precise agreement with the bandwidth obtained from the simulated dispersion.
Accordingly, key-rate optimized entanglement-based quantum key distribution schemes \cite{doi:10.1126/sciadv.1701491} could operate these waveguide sources with up to 26 GHz/mW upon dense wavelength division multiplexing over 70 (100 GHz-spaced) channels. Given this integrated source's potential for unprecedentedly high QKD rates at remarkably low pump powers, further refinement of the coupling interfaces can improve extracted key rates by a manifold. With unequal bandwidth requirements for SHG and SPDC emission signals, a combination of grating (input) and inverted taper-based (output) couplers may be particularly well matched for the short-$\lambda$ TM-mode and the telecom-wavelength TE-Mode, respectively. In this case, the unequal bandwidth requirements accomodate the comparatively relaxed feature size requirements of grating layouts for second and inverse-taper designs for first harmonic coupling. Therefore, with minor improvements on fabrication-technical details, the here presented integrated photonics devices can satisfy practical requirements for widespread QKD applications.

%

\end{document}